\documentclass[letterpaper,twocolumn,10pt]{article}

\usepackage{endnotes}
\usepackage{times}
\usepackage{epsfig}
\usepackage{subfigure}
\usepackage{graphics}
\usepackage{graphicx}
\usepackage{multirow}
\usepackage[latin1]{inputenc}
\usepackage{algorithmic}
\usepackage[plain]{algorithm}

\begin{document}

\date{}


\title{Improving Spam Detection Based on Structural Similarity}

\newcommand{\institutions}{\mbox{{\hspace*{-.5cm}
      \begin{minipage}{7cm}
        \begin{center}
          $^{\dag}$ Computer Science Dept.\\
          Universidade Federal de Minas Gerais \\
          Belo Horizonte - Brazil \\
          \{lhg, fernando, barra, virgilio, jussara\}@dcc.ufmg.br
        \end{center}
      \end{minipage}
      \begin{minipage}{7cm}
        \begin{center}
          $^{\ddag}$ Computer and Computational Sciences\\
          Los Alamos National Laboratory \\
          Los Alamos - USA\\
          lmbett@lanl.gov
        \end{center}
      \end{minipage}}}}

\author{
  {\rm Luiz H. Gomes$^{\dag}$\footnote{Luiz H. Gomes is supported by Banco Central do
      Brasil.}, Fernando D. O. Castro$^{\dag}$, Rodrigo B. Almeida$^{\dag}$, }\\
  {\rm Luis M. A. Bettencourt$^{\ddag}$, Virgílio A. F. Almeida$^{\dag}$, Jussara M. Almeida$^{\dag}$}
\\[0.3cm]
\institutions
}


\maketitle

\thispagestyle{empty}

\subsection*{Abstract}


We propose a new detection algorithm that  uses structural relationships  
between senders and recipients of email as the basis for the identification of
spam messages.  Users and receivers are represented as vectors in their
reciprocal spaces. A measure of similarity  between vectors is constructed and
used to group users into clusters. Knowledge of their classification  as past
senders/receivers of spam or legitimate mail, comming
from an auxiliary detection algorithm, is then used to label these
clusters probabilistically. This knowledge comes from an auxiliary
algorithm. The measure of similarity between the sender and receiver sets of a
new message to the center vector  of clusters is then used to asses the
possibility of that message being legitimate or spam. We show that the proposed
algorithm is able to correct part of the false positives (legitimate messages
classified as spam) using a testbed of one week smtp log. 

\section{Introduction}
\label{sec:introduction}



The relentless rise in spam email traffic,  now accounting for about 
$83\%$ of all incoming messages, up from $24\%$ in January 
2003~\cite{messageLabs},  is becoming one of the greatest threats to the
use of email as a form of communication. 

The greatest problem in detecting spam stems from active adversarial efforts to thwart
classification. Spam senders use a multitude of techniques based on knowledge of
current detection algorithms, to evade detection. These techniques
range from changes in the way text is written - so that it can not be directly 
analyzed computationally,  but can be understood by humans naturally -
to frequent changes in other elements, such as user names, domains, subjects, etc. 
Therefore, good choices for spam identifiers are becoming increasingly more difficult.


In the light of this enormous variability the question then is: what are the
identifiers of spam that are most costly to change, from the point of view of
the sender? The limitations of attempts to recognize spam by analyzing content
are clear~\cite{gerf}.  Content-based techniques\cite{sahami98bayesian,
  zhou03approximate, spamassassin} have to cope with the constant changes in
the way spammers generate their solicitations. The structure of the target
space for these solicitations   tends however  to be much more stable since
spams senders still need to reach recipients, even if under forged identifiers,
in order to be effective. Specifically by structure we mean the space of
recipients targeted by a spam sender, as well as the space of senders that
target a given recipient,  i.e. the contacts of a user. The contact lists, or
subsets thereof, can then be thought of as a signature of spam senders and
recipients. Additionally by    constructing a similarity measure in these
spaces we can track  how lists evolve over time, by addition or removal of
addresses.  
    
In this paper, we propose an algorithm for spam detection that uses structural
relationships between senders and recipients as the basis for the
identification of spam messages.  The algorithm must work in conjunction with
another spam classifier, necessary to produce spam  or legitimate mail tags on
past senders and receivers, which in turn are used to infer new ones through
structural similarity (hereafter called: auxiliary algorithm), 
The key idea is that the lists spammers and legitimate users send messages to,
as well as the lists from which they receive messages from can be used as the 
identifiers of classes of email traffic~\cite{priority, ceas}. 
We will show that the final result of the application of our structural
algorithm over the determinations of  the initial  classifier  leads to the
correction of a number of misclassifications as false positives.  


This paper is organized as follows: Section~\ref{sec:modeling} presents the
methodology used to handle email data. Our structural algorithm is described in
Section~\ref{sec:algorithm}. We present the characteristics of our   
example workload in section~\ref{sec:results}, as well as the classification results 
obtained with our algorithm over this set. Related work is presented in
Section~\ref{sec:related-work} and conclusions and future work in
Section~\ref{sec:concl-future-works}. 
    

\section{Modeling Similarity Among Email Senders and Recipients}
\label{sec:modeling}

Our proposed spam detection algorithm exploits the structural
similarities that exist in groups of senders and recipients 
as well as in the relationship established through the emails
exchanged between them. This section introduces our modeling of
individual email users and a metric to express the similarity existent
 among different users. It then extends the modeling to account for
clusters of users who have great similarity. 

Our basic assumption is that, in both 
legitimate email and spam traffics, users have a defined list of peers 
they often have contact with (i.e., they  send/receive an email 
to/from). In legitimate email traffic, contact lists are consequence of
 social relationships on which users' communications are
based. In spam traffic, on the other hand, the lists used by spammers 
to distribute their solicitations are created for business interest 
and, generally, do not reflect any form of social interaction. 
A user's contact list certainly may change over time. However,
we expect it to be much less variable than other characteristics 
commonly used for spam detection, such as
sender user-name, presence of certain keywords in the email content
and encoding rules. In other words, we expect contact lists to be
more effective in identifying spams and, thus, we use them as
the basis for developing our algorithm.
 
We start by representing an email user as a vector in a 
multi-dimensional conceptual space created with all possible
contacts. We represent email senders and recipients separately. 
We then use vectorial
operations to express the similarity among multiple senders (recipients), 
and use this metric for clustering them.
Note that the term email user is
used throughout this work to denote any identification of
an email sender/recipient (e.g., email address, domain name, etc).

Let $N_r$ be the number of distinct recipients. We represent
sender $s_i$ as a $N_r$ dimensional vector, $\vec{s_i}$, 
defined in the conceptual space created by the email recipients being
considered.  The $n$-th 
dimension (representing recipient $r_n$) of $\vec{s_i}$ is defined as: 
\begin{eqnarray}
   \vec{s_i}[n] = \left\{ \begin{array}{ll}
1, & $ if $s_i \rightarrow r_n \\
0, & $ otherwise$ \\
\end{array} 
\right.,
\end{eqnarray}
where $s_i \rightarrow r_n$ indicates that sender $s_i$ has sent at least one email to
$r_n$ recipient. 

Similarly, we define $\vec{r_i}$ as a $N_s$ dimensional vector representation
for the recipient $r_i$, where $N_s$ is the number of distinct senders being
considered. The $n$-th dimension of this vector is set to $1$ if recipient
$r_i$ has received at least one email from $s_n$.


We next define the similarity between two senders $s_i$ and $s_j$ 
as the cosine of the angle between their vector representation 
($\vec{s_i}$ and $\vec{s_j}$). The similarity is computed as follows:
\begin{eqnarray} \label{similarity}
  sim(s_i,s_j) = \frac{\vec{s_i} \circ \vec{s_j}}{|\vec{s_i}||\vec{s_j}|} = cos(\vec{s_i},\vec{s_j}) ,
\end{eqnarray}
where $\vec{s_i} \circ \vec{s_j}$ is the internal product of the vectors and
$|\vec{s_i}|$ is  the norm of $\vec{s_i}$. Note that this 
metric varies from 0, when senders do not share any recipient in their 
contact lists,  to 1, when senders have identical contact lists and thus
have the same representation. The similarity between two recipients 
is defined similarly.

We note that our similarity metric has different interpretations 
in legitimate
and spam traffics. In legitimate email traffic, it represents
 social interaction with the same group of people, whereas in the spam
traffic, a great similarity represents the use of different identifiers by
the same spammer or the sharing of distribution lists by distinct spammers.

Finally, we can use our vectorial modeling approach to represent a 
cluster of users (senders or recipients) who have great similarity.
A sender cluster $sc_i$, represented by vector
 $\vec{sc_i}$, is computed as the vectorial sum of its elements, 
that is:   
\begin{equation}
  \vec{sc_i} = \sum_{s \in sc_i}{\vec{s}}.
\end{equation}

The similarity between 
sender $s_i$ and an existing cluster $sc_j$ can then be directly 
assessed by extending Equation~\ref{similarity} as follows:
\begin{eqnarray}
   sim(sc_i,s_i) = \left\{ \begin{array}{ll}
 cos(\vec{sc_i} - \vec{s_i}, \vec{s_i}) , & $ if $s_i \in sc_i \\
cos(\vec{sc_i}, \vec{s_i}) , & $ otherwise$ \\
\end{array} 
\right.
\end{eqnarray}
We note that a sender $s_i$  vectorial representation and thus the sender cluster to which
it belongs (i.e., shares the greatest similarity)  may change over time
as new emails are considered. Therefore, in order to accurately estimate 
 the similarity
between a sender $s_i$ and a sender cluster $sc_i$ to which $s_i$ currently belongs, we 
first remove $s_i$ from $sc_i$, and then take the cossine between the two vectors
 ($\vec{sc_i} - \vec{s_i}$ and $\vec{s_i}$). This is performed so that
the previous classification of a user does 
not influence its reclassification. Recipient clusters and the similarity 
between a recipient and a given recipient cluster are defined analogously.

  
\section{A New Algorithm for Improving Spam Detection }
\label{sec:algorithm}

This section introduces our new email classification 
algorithm which exploits the similarities between 
email senders and between email recipients for 
clustering and uses historical properties of clusters 
to improve spam detection accuracy. Our algorithm is 
designed to work together with any existing spamdetection 
and filtering technique that runs at the ISP level. 
Our goal is to provide a significant reduction of 
false positives (i.e., legitimate emails wrongly classified as spam), 
which can be as high as 15\% in current filters~\cite{sizecost}.

A description of the proposed algorithm is shown 
in Algorithm~\ref{alg:detection}. 
It runs on each arriving email $m$, taking as input 
the classification of $m$, $mClass$, as either spam 
or legitimate email, performed by the existing auxiliary 
spam detection method. Using the vectorial representation 
of email senders, recipients and clusters as well as the 
similarity metric defined in Section 2, it then determines a 
new classification for $m$, which may or not agree with $mClass$. 
The idea is that the classification by the auxiliary method 
is used to build an incremental historical knowledge base that gets 
more  representative through time. Our algorithm benefits from that and 
outperforms the  auxiliary one as shown in Section~\ref{sec:results}. 

\begin{algorithm}  
\centering  
    \begin{algorithmic}    
        \FORALL{arriving message $m$}      
            \STATE $mClass = $classification of $m$ by auxiliary detection method;      
            \STATE $sc = $find cluster for $m.sender$;      
            \STATE Update spam probability for $sc$ using $mClass$;      
            \STATE $P_s(m) = $spam probability for $sc$;      
            \STATE $P_r(m) = 0$;      
            \FORALL{recipient $r \in m.recipients$}        
                \STATE $rc = $find cluster for $r$;        
                \STATE Update spam probability for $rc$ using $mClass$;        
                \STATE $P_r(m) = P_r(m) + $spam probability for $rc$;      
            \ENDFOR      
            \STATE $P_r(m) = P_r(m)/size(m.recipients)$      
            \STATE $SP(m) = $ compute spam rank based on $P_s(m)$ and  $P_r(m)$;      
            \IF{$SP(m) > \omega$}        
                \STATE classify $m$ as spam;      
            \ELSIF{$SP(m) < 1 - \omega$}        
                \STATE classify $m$ as legitimate;      
            \ELSE        
                \STATE classify $m$ as $mClass$;      
            \ENDIF     
        \ENDFOR  
    \end{algorithmic}  
\caption{New Algorithm for  Email Classification}   
\label{alg:detection}
\end{algorithm}

In order to improve the accuracy of email classification, our algorithm 
maintains sets of sender and recipient clusters, created based on the structural similarity
of different users. A sender (recipient) of an incoming email
 is added to a sender (recipient) cluster
that is most similar to it, as defined in Equation (4), provided that their similarity 
exceeds a given threshold $\tau$. Thus, $\tau$ defines the minimum similarity
a sender (recipient) must have with a cluster to be assigned to it. 
Varying $\tau$ allows us to create more tightly or loosely knit
clusters. If no cluster can be found, a new single-user cluster is created. 
In this case, the sender (recipient) is used as  seed for populating the new 
cluster.   

The sets of recipient and sender clusters are updated at each new email arrival 
based on the email sender and list of recipients. Recall that to determine the
cluster a previously observed, and thus clustered, user (sender or recipient) 
belongs to, we first remove the user from his current cluster and then assess 
its similarity to each existing cluster. Thus, single-user clusters tend to 
disappear as more emails are processed, except for users that appear only very sporadically. 

\begin{figure}[th!]
  \centering
  \includegraphics[width=200pt]{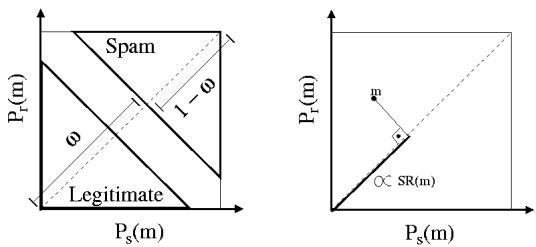}
  \caption{Spam Rank Computation and Email Classification.}
  \label{fig:spamRank}
\end{figure}

A probability of sending (receiving) a spam is assigned to each sender (recipient)
 cluster. We refer to this measure as simply the cluster spam probability. 
We calculate the spam probability of a sender (recipient)
cluster as the average spam probability  of its elements, which, in 
turn, is estimated based on the frequency of spams sent/received by each of them 
in the past. Therefore, our algorithm uses the result of the email classification performed 
by the auxiliary algorithm on each arriving email $m$ ($mClass$ in 
Algorithm~\ref{alg:detection}) to continuously update cluster spam probabilities.

Let us define the probability of an email $m$ being sent by a spammer, 
$P_s(m)$, as the spam probability of its sender's cluster. 
Similarly, let the probability of an email
 $m$ being addressed to users that receive spam, $P_r(m)$, as the
average spam probability of all of its recipients' clusters 
(see Algorithm~\ref{alg:detection}). Our algorithm uses 
$P_s(m)$ and $P_r(m)$ to compute a number that
expresses the chance of email $m$ being spam. We call this number the spam rank of
email $m$, denoted by $SR(m)$. The idea is that emails with large values of $P_s(m)$
and $P_r(m)$ should have large spam ranks and thus should be classified as spams.
Similarly, emails with small values of $P_s(m)$
and $P_r(m)$ should receive low spam rank and be classified as legitimate email. 

Figure~\ref{fig:spamRank} shows a graphical representation of
the computation of an email spam rank. We first normalize the probabilities
$P_s(m)$ and $P_r(m)$  by a factor of $\sqrt{2}$, so that the diagonal of
the square region defined in the bi-dimensional space  is equal to 1 
(see Figure~\ref{fig:spamRank}-left). Each email $m$ can be represented as
 a point in this square. The spam rank of $m$, $SR(m)$, is 
then defined as the  length  of the segment starting at
the origin (0,0) and ending at the projection of $m$
 on the diagonal of the square (see Figure~\ref{fig:spamRank}-right). Note
the spam rank varies between 0 and 1.

The spam rank $SR(m)$ is then used to classify $m$ as follows: if it is greater
than a given threshold $\omega$, the email is classified as spam; if it is 
smaller than $1 - \omega$, it is classified as legitimate email. Otherwise, we can
not precisely classify the email, and we rely on the initial classification
provided by the auxiliary detection algorithm. The parameter $\omega$ can be tuned
to determine the precision that we expect from our classification. 
Graphically, emails are classified according to the marked regions shown in
Figure~\ref{fig:spamRank}-left. The two triangles, with identical size 
and height $\omega$, represent the regions where 
our algorithm is able to classify emails as either spam (upper right) or legitimate
email (lower left). 


\section{Experimental Results}
\label{sec:results}

In this section we describe our experimental results. We first present some important details of 
our workload, followed by the quantitative results of our approach, compared to others.

\subsection{Workload}
\label{sec:workload}
                                                           
Our email workload consists of anonymized and sanitized SMTP logs of incoming
emails to a large university in Brazil, with around 22 thousand students. The
server  handles all emails coming from domains
outside the university, sent to students, faculty and staff with
email addresses under the university's domain name~\footnote{Only the
emails addressed to two out of over 100 university subdomains (i.e.,
departments, research labs, research groups) do not pass through the central
server.}
      
The central email server runs Exim email software~\cite{exim},
the Amavis virus scanner~\cite{amavis} and the Trendmicro
Vscan anti-virus tool~\cite{antivirus}. A set of pre-acceptance
spam filters (e.g. black lists, DNS reversal) blocks about 50\% of the total traffic
received by the server.   

The messages not rejected by the pre-acceptance  tests are directed to
Spam-Assassin~\cite{spamassassin}. Spam-Assassin is a popular spam filtering
software that detects spam messages  based on a changing set of user-defined
rules. These rules assign scores to each email received based on the presence
in the  subject or in the email body of one or more pre-categorized
keywords. Spam-Assassin also uses other rules based on message size 
and encoding. Highly ranked messages according to these criteria are flagged 
as spam. 

We analyze an eight-day log collected  between 01/19/2004 to 01/26/2004. 
Our logs store the header of each email (i.e. containing sender, recipients, size , date, etc.) 
that passes the pre-acceptance filters, along with the results  of the tests performed by
Spam-Assassin and the virus scanners. We also have the full body of the messages
that were classified as spam by Spam-Assassin. Table~\ref{bst} summarizes our
workload.

\begin{table}[th!]
  \centering
  \footnotesize
  \begin{tabular}{|l|l|l|l|} \hline 
    {\bf Measure}   & {\bf Non-Spam}  & {\bf Spam} &  {\bf Aggregate} \\ \hline  
    \# of emails & 191,417 & 173,584& 365,001    \\ \hline
    Size of emails& 11.3 GB& 1.2 GB& 12.5 GB \\ \hline
    \# of distinct senders & 12,338& 19,567& 27,734 \\ \hline
    \# of distinct recipients & 22,762& 27,926& 38,875\\ \hline
  \end{tabular}
  \caption{Summary of the Workload}
  \label{bst}
\end{table}

By visually inspecting the list of sender {\em user names}~\footnote{The part
  before @ in email addresses.} in the  spam component of our workload, we
found that a large number of them corresponded to a seemingly random sequence
of characters, suggesting that spammers tend to change user names as an evasion
technique. Therefore, for the experiments presented below we identified the
sender of a message by his/her domain while recipients were identified by their
full address, including both domain and user name. 

\subsection{Classification Results}
\begin{figure}[th!]
  \centering
  \includegraphics[width=200pt]{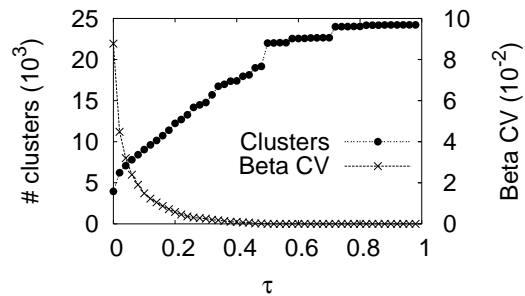}
  \caption{Number of Email User Clusters and Beta CV  vs. $\tau$.} 
  \label{fig:betacvxncom}
\end{figure}

The results shown in this section were obtained through the simulation of
the algorithm proposed here over the set of messages in our logs. The
implementation of the simulator made use of an inverted 
lists~\cite{moffat} approach for storing information about senders, recipients
and clusters that is effective both in terms of memory and processing
time. Our simulations were executed on a commodity workstation
(Intel Pentium \textregistered 4 - 2.80GHz - with 500MBytes) and
the simulator was able to classify 20 messages per second. This is far faster
than the average rate with which messages usually arrive and than the peak rate
observed over the workload collection time~\cite{gomes}.    


\begin{figure}[th!]
  \centering
  \subfigure[Bin size = 0.10]{\includegraphics[width=100pt]{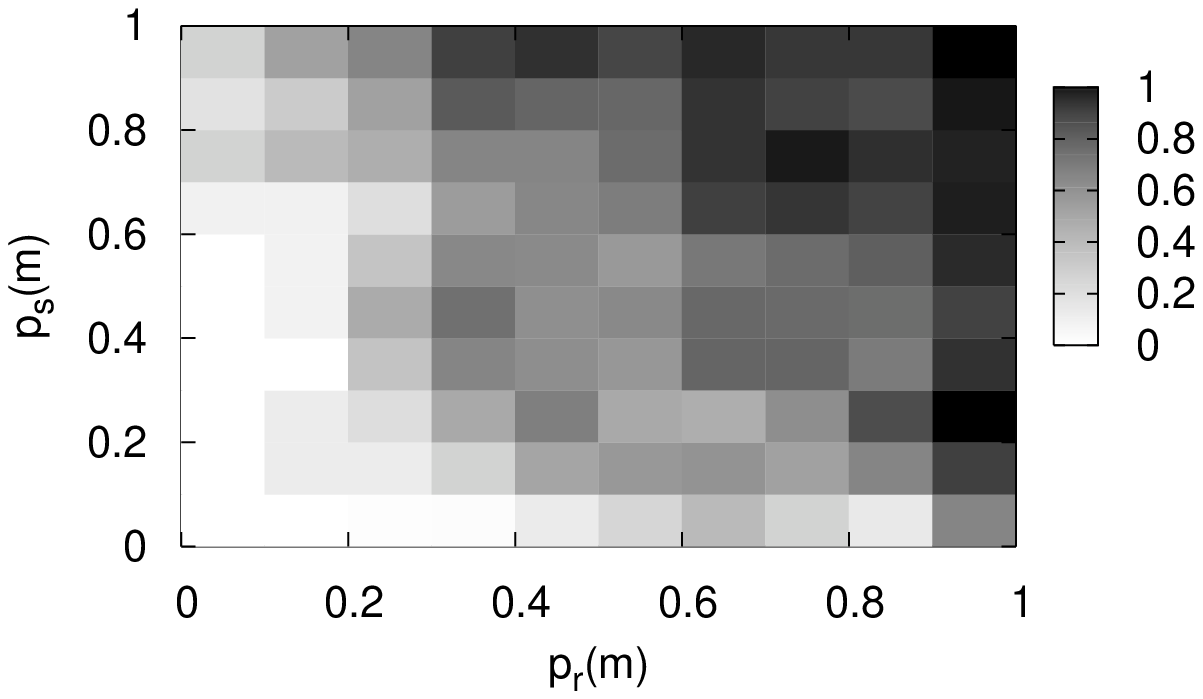}}
  \subfigure[Bin size = 0.25]{\includegraphics[width=100pt]{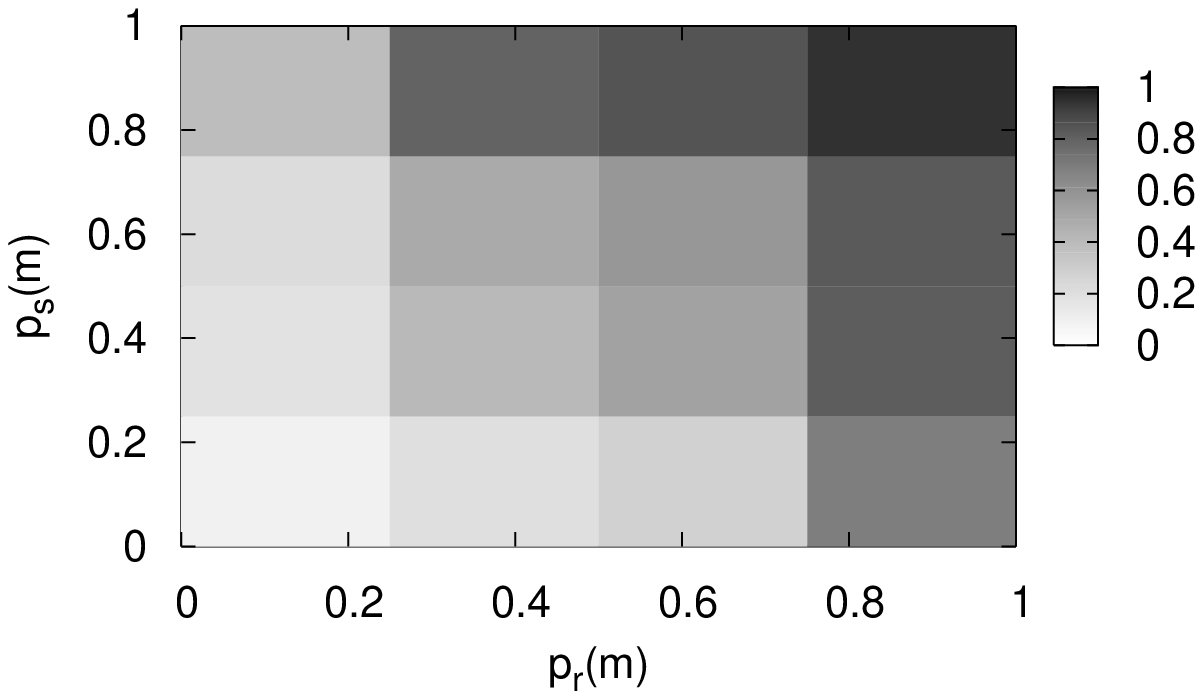}}
  \caption{Number of Spam Messages by Varying Message Spam Probabilities for
    Different Bin Sizes.}
  \label{fig:message_classification}
\end{figure}



The number and quality of the clusters generated through our similarity
measure are the direct result of the chosen value for the threshold $\tau$ (see
Section~\ref{sec:algorithm}). In order to determine the best parameter value
the simulation was executed several times for varying $\tau$. 

Figure~\ref{fig:betacvxncom} shows how the number of clusters
and beta CV~\footnote{Beta CV means intra CV/inter
      CV and assesses the quality of the clusters generated. The lower the beta CV
      the better quality in terms of grouping obtained~\cite{livrovirgilio}.}
    vary with $\tau$. There is 
one clear  point of stabilization of the curve (i.e. a plateau) at $\tau = 0.5$
and that is the value we adopt for the remaining  of the paper. Although other 
stabilization points occur for values of $\tau$ above $0.5$, the lowest of such
values seems to be the most appropriate for our experiments. The reason for
that is that this value of $\tau$ is the one that generates the smaller stable number
of clusters, i.e. cluster with more elements, and that allows us to evaluate
better the beneficial effects that clustering senders and recipients may have.
Moreover, while analyzing the beta CV we are able to see that the quality of
the clustering for all values  $\tau>0.4$ is approximately the same. 

One of the hypothesis of our algorithm is that we can group spam messages in
terms of the probabilities $P_s(m)$ and
$P_r(m)$. Figure~\ref{fig:message_classification} shows the fraction of spam 
messages that exist for different values of $P_s(m)$ and $P_r(m)$ grouped based 
on a discretization of the full space represented in the plot. The full space
is subdivided  into smaller squares of the same size called bins. 
Clearly, spam/legitimate messages are indeed located in the regions (top and bottom 
respectively) as we have hypothesized in Section~\ref{sec:algorithm}. There is however a region in the middle where
we can not determine the classification for the messages based on the computed
probabilities. This is why it becomes necessary to vary $\omega$. One should
adjust $\omega$ based on the level of confidence he/she has on the auxiliary algorithm. 

Figure~\ref{fig:message_classification} shows that differentiation between
senders and recipients for detecting spam can be more effective than the
simple choice we use in this paper. Messages addressed to recipients that 
have high $P_r(m)$ tend to be spam more frequently than messages with the same 
value of $P_s(m)$. Analogously, messages with low $P_s(m)$ have higher probability
of being legitimate messages. Ways of using this information in our algorithm are 
an ongoing research effort that we intend to pursue in future extensions.

\begin{figure}[th]
  \centering
  \includegraphics[width=150pt]{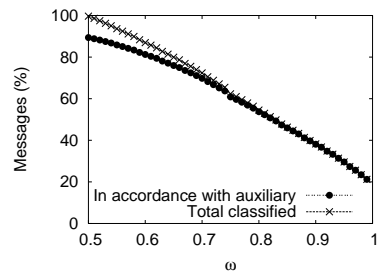}
  \caption{Messages Classified in Accordance With to the Auxiliary Algorithm and the Total
  Number of Messages Classified by Varying $\omega$}
  \label{fig:omega}
\end{figure}

Our algorithm makes use of an auxiliary spam detection algorithm - such as
SpamAssassin. Therefore, we need to evaluate how frequently we maintain the
same classification as such an algorithm. Figure~\ref{fig:omega} shows the
the percentage of messages that received the same classification and the total
number of classified messages in our simulation by varying $\omega$. The
difference between these curves is the set of messages that 
were classified differently from the original classification provided. 
There is a clear tradeoff between the total number of messages that
are classifiable and the accordance with the previous classification provided
by the original classifier algorithm.


In another experiment, we simulated a different algorithm that also makes 
use of history information provided by an auxiliary spam detector described in~\cite{priority}. 
This approach tries to classify messages based on the
historical properties of their senders. We built a simulator for this algorithm
and executed it against our data set. The results show that it was
able to classify $85.11\%$ of the messages in accordance 
with the auxiliary algorithm. Its important to note that, on the other hand,
our algorithm can be tuned by the proper set of threshold $\omega$. The higher
the parameter $\omega$ the more in acordance with the auxiliary classification
the classification of our algorithm is.

We believe that the differences between the original classification and the
classification proposed for high $\omega$ values generally are due to
missclassifications by the auxiliary algorithm. In our data set we have access to
the full body of the messages that were originally classified as spam. Therefore, we
can evaluate a fraction of the total amount of false positives (messages that
the auxiliary algorithm classify as spam and our algorithm  classify as
legitimate message) that were generated by the auxiliary algorithm. This is  
important since there is a common belief that the cost of false positives is
higher than the cost of false negatives~\cite{gerf}.  

Each of the possible false positives were manually evaluated by three people so
as to determine whether such a message was indeed spam. Table~\ref{tab:manual}
summarizes the results for $\omega = 0.85$, 879 messages
were manually analyzed ($0.24\%$ of the total of messages). Our algorithm outperforms the
original classification since it generates less false positives. We emphasize
that we can not similarly determine the quality of classification for the
messages classified as legitimate by the auxiliary algorithm since we do not have
access to the full body of those messages. Due to the cost of manually classifying
messages we can not aford to classify all of the messages classified as spam by
the auxiliary algorithm.
                               
\begin{table}[th!]
  \centering
  \footnotesize
  \begin{tabular}{|l|c|} \hline 
    \multicolumn{1}{|c|}{\bf Algorithm}   & \multicolumn{1}{c|}{\bf \% of Missclassifications } \\ \hline  
    Original Classification & $60.33\%$ \\ \hline
    Our approach & $39.67\%$ \\ \hline
  \end{tabular}
  \caption{Possible False Positives Generated by the Approaches Studied.}
  \label{tab:manual}
\end{table}


\section{Related Work}
\label{sec:related-work}

Previous work have focused on reducing the impact of spam.
The approaches to reduce spam  can be categorized into pre-acceptance and 
post-acceptance methods, based on whether they detect and block
spam before or after accepting messages. Examples of pre-acceptance methods
are black lists~\cite{blacklist2}, gray lists~\cite{greylist}, server
authentication~\cite{spam,authentication} and
accountability~\cite{solvingspam}. Post-acceptance methods are mostly based on
information available in the body of the messages and include Bayesian
filters~\cite{sahami98bayesian}, collaborative
filtering~\cite{zhou03approximate}.

Recent papers have focused on spam combat techniques based 
on characteristics of graph models of email
traffic~\cite{emailnetcombat,spammachines}. The techniques 
used try to model
email traffic as a graph and detect spam and spam attacks 
respectively in terms
of graph properties. In~\cite{emailnetcombat} a graph is
created representing the email traffic captured in the mailbox of individual
users. The subsequent analysis is based on the fact that such a
network possesses several disconnected components. The clustering coefficient
of each of these components is then used to characterize messages as spam or
legitimate. Their results show that 53\% of the messages were 
precisely classified using
the proposed approach. 
In~\cite{spammachines} the authors used the approach of 
detecting machines that 
behave as spam senders by analyzing a border flow graph of sender and recipient
machines. In\cite{priority}, the authors propose a new scheme for
handling spam. It is a post-acceptance mechanism that processes
mail suspected of being spam at reduced priority, when compared to 
the priority assigned to messages classified as legitimate. The 
proposed mechanism\cite{priority} works in conjunction with some sort
of mail filter that provides past history of mails received by a server. 

None of the existing spam filtering mechanisms are 
infallible\cite{priority, gerf}. Their main problems are false positive and
wrong mail classification. In addition to those problems, filters must
be continuously updated to capture the multitude of mechanism constantly
introduced by spammers to avoid filtering actions. The algorithm presented in
this paper aims at improving the effectiveness of spam filtering mechanisms, 
by reducing false positives and by providing information that help those mechanism 
to tune their collection of rules.


\section{Conclusions and Future Work}
\label{sec:concl-future-works}

In this paper we proposed a new spam detection algorithm based on  the structural similarity between contact lists of email users. The idea is that contact lists, integrated 
over a suitable amount of time, are much more stable identifiers of email users than id names, domains
or message contents, which can all be made to vary quickly and widely.
The major drawback of our approach is that our algorithm can only group users based on their structural 
similarity, but has no way of determining by itself if such vector clusters correspond to spam or legitimate email. Because of this feature it must work in tandem with an original classifier.
Given this information we have shown that we can successfully group spam and legitimate email users separately and that this structural inference can improve the quality of other spam detection algorithms.

Specifically we have implemented a simulator based on data collected from the
main SMTP server for a major university in Brazil that uses SpamAssassin. We
have shown that our algorithm can be tuned to produce classifications similar
to those of the original classifier algorithm and that, for a certain set of
parameters, is was capable of correcting false positives generated by
SpamAssassin in our workload.

 There are several improvements and developments that were not explored here, but promise 
 to reinforce the strength of our approach. We intend to explore these in future work. We observe that structural similarity gives us a basis for time correlation of similar addresses, and as such to follow the time evolution of spam sender techniques, in ways that suitably factor out the enormous variability of their apparent identifiers. Finally we note  that the probabilistic basis of our approach lends itself naturally to the evolution of users' classifications (say through Bayesian inference), both through collaborative filtering using user feedback and from information derived from other algorithmic classifiers. 
 

\bibliographystyle{acm}

\end{document}